\documentclass[12pt]{article}
\usepackage{graphicx}
\usepackage{pslatex}

\author{V. Parihar, A. Widom, Y. N. Srivastava$\dagger$\\
        Physics Department, Northeastern University, Boston MA USA\\
        and\\ 
        $\dagger$ Physics Department 
        \& INFN, University of Perugia, Perugia IT
}

\title{Dynamic Time Scales in Colored Glass Nuclear Matter}
\date{}
\begin{document}

\maketitle

\pagestyle{headings}

\section{Introduction \label{Intro}}

Ultra high energy collisions between heavy nuclei can be employed to experimentally 
probe the QCD properties of dense quark-gluon 
systems \cite{Maiani:2005,Shuryak:2004,Shuryak:2002,Soff:2001,Teaney:2001}. 
The properties of the high density quark gluon system have also been compared to a color 
glass\cite{Gtulassy,Kharzeev,Iancu,Jalilian,Venugopalan,
Iancu:02,Venugopalan:97,Ayala}.
High energy proton-proton scattering also probes the QCD vacuum. 
The QCD vacuum differs from the QED vacuum 
via properties largely due to the nature of the electrically charged and color charged 
dielectric screening response functions. In the QCD case, there is a quark-quark 
potential containing both a Coulomb-like piece as well as a linear {\em confining} 
potential. The confining  {\em force} may be viewed in terms of the tension 
\begin{math} \sigma  \end{math} in a ``string'' connecting two quarks having the 
experimental value 
\begin{equation}
\sigma \approx \frac{0.18\ {\rm GeV}^2}{\hbar c}
\approx 1.45\times 10^5\ {\rm Newton}.
\label{intro1}
\end{equation}
The QCD string may be visualized as a color electric field flux tube. The string 
vibrational and rotational states describe the mesons contributing to nuclear forces. 
The QCD string has a much lower tension than some theoretical strings said to 
describe gravity; e.g. 
\begin{equation}
\sigma_{\rm gravity}=\frac{c^4}{8\pi G}\approx 4.8\times 10^{42}\ {\rm Newton}.
\label{intro1a}
\end{equation}
In what follows only properties of the low tension QCD string will be invoked.

As the string excitation energies grow ever larger, the enormous number of possible string 
excitations can be described by the entropy as a function of energy. In computing 
quantum transitions, the number \begin{math} \Omega  \end{math} of possible quantum states  
is determined by the entropy \begin{math} S \end{math} via 
\begin{equation}
S=k_B \ln \Omega .
\label{intro2}
\end{equation}
Since one must average over initial states and sum over the final states, transition 
rates have the ``detailed balance'' form 
\begin{equation}
\Gamma_{i\to f}=\nu_\infty \left\{\frac{\Omega_f}{\Omega_i} \right\}
=\nu_\infty \exp\left[\frac{S_f-S_i}{k_B}\right].
\label{intro3}
\end{equation}
Eq.(\ref{intro3}) will be applied to the time scales of nuclear matter motions which occur 
shortly after an ultra high energy collision between two heavy nuclei. The time scales are 
familiar from the the theory of 
glasses\cite{Vogel:1921, Fulcher:1925, Tammann:1926, Tanaka:2003}. 
The time scales are\cite{vivek:2006} 
\begin{equation}
\tau=\tilde{\tau }\ \exp\left[\frac{\Phi }{k_B(T-T_H)}\right] ,
\ \ \ {\rm for}\ \ \ T>T_H ,
\label{intro4}
\end{equation}
wherein \begin{math} T_H \end{math} is the Hagedorn temperature\cite{Hagedorn:1965} 
and \begin{math} \Phi \end{math} is the activation energy. Experimentally, 
\begin{equation}
k_BT_H \approx 207\ {\rm MeV}\ \ \ {\rm and}
\ \ \ \Phi \equiv \frac{7k_BT_H}{2} \approx 725\ {\rm MeV}.
\label{intro5}
\end{equation}
The notion of final state particle product entropy is also of use 
in discussing the total cross section as a function of center of mass energy. 

\section{Vacuum Stability and Dielectric Screening \label{VSDS}}

The Vacuum law of force between two charges $z_1e$ and $z_2e$ may be written as 
\begin{equation}
V(r)=e^2z_1z_2\int \left[\frac{e^{i{\bf Q\cdot r}}}{Q^2\varepsilon (Q^2)}\right]
\frac{d^3{\bf Q}}{(2\pi )^3}=\left[\frac{z_1z_2e^2}{4\pi r}\right]\chi(r),
\label{VSDS1}
\end{equation}
wherein the vacuum screening function $\chi (r)$ is determined by the vacuum 
dielectric response function $\varepsilon (Q^2)$ via 
\begin{equation}
\chi(r)=\frac{2}{\pi }\int_0^\infty \sin(Qr)
\left[\frac{dQ}{Q\varepsilon(Q^2)}\right].
\label{VSDS2}
\end{equation}
In the more usual notation of QED, one defines the dielectric response of the 
vacuum in terms of a running coupling strength 
\begin{equation}
\alpha=\frac{e^2}{4\pi \hbar c}\ \ \ \Rightarrow 
\ \ \ \alpha(Q^2)=\frac{e^2}{4\pi \hbar c \varepsilon (Q^2)}\ .
\label{VSDS3}
\end{equation}
In the limit of large distances, $\ \hbar /mc \ll r\ $, we have 
\begin{equation}
\lim_{Q^2\to 0^+}\varepsilon(Q^2)=\lim_{r\to \infty}\chi(r)=1,
\label{VSDS4}
\end{equation}
which implies a simple Coulomb law. The subtracted dispersion relation 
then reads 
\begin{equation}
\varepsilon(Q^2)=1-\frac{Q^2}{\pi }\int_0^\infty 
\left[\frac{{\Im}m\ \varepsilon(-\nu -i\ 0^+)}{\nu +Q^2}\right]
\frac{d\nu }{\nu }\ .
\label{VSDS5}
\end{equation}
One has a dissipative QED vacuum with positive electrical conductivity 
for time-like wave vectors, 
\begin{equation}
{\Im}m\ \varepsilon(-\nu -i\ 0^+)\ge 0,
\label{VSDS6}
\end{equation}
at the expense of a Landau $\alpha (Q^2)$ singularity for some 
space-like wave vector implicit in Eqs.(\ref{VSDS3}) and (\ref{VSDS6}).  
Such a singularity has been referred to as a Landau 
ghost\cite{Landau:1955, siva, Berestetskii:1997, yogi}.

For the QCD case the potential between color charges analogous to 
Eq.(\ref{VSDS1}) may be written as 
\begin{eqnarray}
V_s(r)=\eta^{ab}t_{1a}t_{2b}\left(\frac{g^2}{4\pi r}\right)\chi_s(r),
\nonumber \\ 
\chi_s(r)=\frac{2}{\pi }\int_0^\infty \sin(Qr)
\left[\frac{dQ}{Q\varepsilon_s(Q^2)}\right],
\label{VSDS7}
\end{eqnarray}
wherein the strong screening dielectric response arises from the 
strong running coupling constant 
\begin{equation}
\alpha_s=\frac{g^2}{4\pi \hbar c}\ \ \ \Rightarrow 
\ \ \ \alpha_s(Q^2)=\frac{g^2}{4\pi \hbar c \varepsilon_s (Q^2)}\ .
\label{VSDS8}
\end{equation}
The subtracted dispersion relation for the strong color screening 
response function is governed by 
\begin{eqnarray}
\varepsilon_s(Q^2)=-\frac{Q^2}{\pi }\int_0^\infty 
\left[\frac{{\Im}m\ \varepsilon_s(-\nu -i\ 0^+)}{\nu +Q^2}\right]
\frac{d\nu }{\nu }\ ,
\nonumber \\ 
-\frac{{\Im}m\ \varepsilon_s(-\nu -i\ 0^+)}{\pi}=\frac{\alpha_s}{4\pi }
\left[\frac{11}{3}N_c-\frac{2}{3}n_f\right]+\ldots\ .
\label{VSDS9}
\end{eqnarray}
 For QCD\cite{pacetti} it then follows that 
 $\varepsilon_s(Q^2)\ge 0$ so there is no QCD Landau ghost. 
 On the other hand, the color 
 conductivity is {\em not} dissipative. In detail,
 \begin{equation}
 {\Im m}\ \varepsilon_s(-\nu -i0^+)\le 0
 \ \ \ \ \ \ {\rm (color\ amplifier)}.
 \label{VSDS10}
 \end{equation}
The {\em perturbative} QCD vacuum as described by Eqs.(\ref{VSDS9}) and 
(\ref{VSDS10}) is thereby unstable. The length scale $L$ associated with 
strong color screening $\varepsilon_s(Q^2)$, i.e. 
\begin{equation}
L^2=-2\lim_{Q^2\to 0^+}\frac{\varepsilon_s(Q^2)}{Q^2}\ .
\label{VSDS11}
\end{equation}
The two body color potential (as \begin{math} r\to \infty \end{math}) 
may then be written as a QCD string linear potential 
\begin{equation}
V_s(r)=-\eta^{ab}t_{1a}t_{1b} \sigma\ r  \ \ \ \ \ {\rm where}
\ \ \ \ \ \sigma =\frac{g^2}{4\pi L^2} \ .
\label{labelVSDS12}
\end{equation}
is the string tension.

\section{Nuclear Matter and String Models \label{NMSM}}

The liquid droplet model of a nucleus, assumes a 
{\em non-relativistic dilute fluid} of atomic mass   
\begin{math} A \end{math} and radius 
\begin{math} R=aA^{1/3}\approx 
1.2\times 10^{-13}{\rm cm}\ A^{1/3}\end{math}
filling up a Fermi sphere of momentum radius 
\begin{math} p_F=\hbar k_F \end{math}, 
\begin{equation}
A=\left[\frac{4\pi R^3}{3}\right]\left[\frac{4}{(2\pi)^2}\right] 
\left[\frac{4\pi k_F^3}{3}\right]=\frac{8}{9\pi }(k_FR)^3.
\label{NMSM1}
\end{equation}
Eq.(\ref{NMSM1}) allows for the computation of the Fermi velocity 
\begin{math} v_F=\hbar k_F/M  \end{math}; i.e. 
\begin{equation}
\frac{v_F}{c}=\left(\frac{9\pi}{8}\right)^{1/3}
\frac{\hbar}{Mca}\approx \frac{1}{4}.
\label{NMSM2}
\end{equation}
The liquid drop thereby appears to be internally relativistic. 
Another notion of nuclear liquids being relativistic arises 
in shell models. A strong \begin{math} {\bf L\cdot S}  \end{math} 
coupling exists characteristic of the Dirac relativistic spinor 
model of fermions. The QCD string constituents of nuclear matter 
starts from a relativistic viewpoint.  

In the string model, the atomic number 
\begin{math} Z=2N_u+N_d \end{math} 
and the atomic mass 
\begin{math} A=3(N_u+N_d) \end{math}
can be expressed in terms of the net numbers of 
up and down quarks. The mesons internal to the 
nucleus are made of quark anti-quark pairs 
connected by strings. The strings consist of color 
electric flux tubes. The electric field colorless 
condensate \begin{math} 
{\cal E}=
\sqrt{\eta_{ab}\left<{\bf E}^a \cdot {\bf E}^b
-{\bf B}^a \cdot {\bf B}^b\right>}
\end{math}     
within the string determines the the magnitude of the 
string tension \begin{math} \sigma =g{\cal E} \end{math}.

To understand the spectrum of excited states of a QCD string, 
one may consider a length \begin{math} \Lambda \end{math} of a 
rotating string along with an eigenvalue angular velocity 
\begin{equation}
\omega =\frac{\pi c}{\Lambda }=\frac{\pi c \sigma }{E}
=\frac{\pi  \sigma }{Mc}
\label{NMSM3}
\end{equation} 
wherein \begin{math} E=Mc^2=\sigma \Lambda  \end{math} 
is the energy of the string. Since the angular velocity 
of the string with angular momentum 
\begin{math} J \end{math} obeys 
\begin{equation}
\omega =\frac{dE}{dJ}=\frac{\pi c \sigma }{E}
\ \ \ \ \ \ \Rightarrow 
\ \ \ \ \ \ \frac{dJ}{d(E^2)}=\frac{1}{2\pi c\sigma }\ ,
\label{NMSM4}
\end{equation} 
it follows that the angular momentum energy-squared variation 
is a linear classical Regge trajectory.
The quantum mechanical version of Eq.(\ref{NMSM4}) is usually written 
\begin{equation}
J=J_0+\left(\frac{E^2}{2\pi c\sigma}\right)
=\hbar(\alpha_0+\alpha^\prime M^2)
=\hbar \alpha_0+\left(\frac{c^3}{2\pi \sigma }\right)M^2.
\label{NMSM6}
\end{equation}
From the experimental slope of the meson spectrum, the measured 
value of the QCD string tension is given in Eq.(\ref{intro1}). As the energy 
grows ever larger, the degeneracy of the excited states grows 
exponentially. The resulting increasing entropy and the large 
string energy degeneracy leads to notion of a Hagedorn temperature.    

\section{String Entropy \label{SE}}

\begin{figure}[tp]
\scalebox {0.6}{\includegraphics{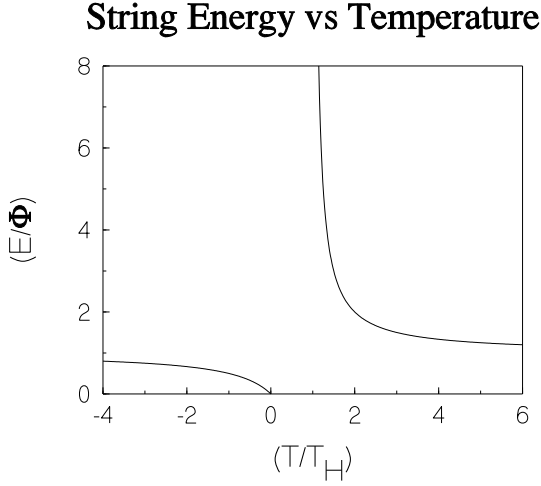}}
\caption{The temperature in the interval $0<T<T_H$ is forbidden. 
For a small amount of energy $E<\Phi$ fed into a string from a 
high energy nuclear collision, the temperature will be negative. 
In the high energy transfer regime, $E>\Phi $, the temperature will be  
positive. The temperature passes from negative to positive through 
the points $T=\pm \infty$ as $E$ passes through $\Phi$. For the very highest 
collision energy passed to a string, the temperature will ``cool'' down 
from above to $T_H$. The Hagedorn temperature is the lowest possible positive  
temperature of the string. In this regard, note the negative heat capacity 
$C=(dE/dT)<0$.}
\label{fig1}
\end{figure}

For a fixed angular momentum $J=\hbar N$ and for two polarizations 
\begin{equation}
N=\sum_{p=1,2}\ \sum_{k=1}^\infty kN_{kj}.
\label{SE1}
\end{equation}
The number of partitions of the integer $N$ into the sum of other integers equal 
or less than $N$ yields the degeneracy number $\Omega $\cite{Hardy:1918, Huang:1970}. 
For large $N$,
\begin{eqnarray}
\ln \Omega(N)&=&2\pi \sqrt{\frac{2N}{3}}
+\ln\left(\frac{3^{1/2}}{(6N)^{7/4}}\right)+\ldots \ ,
\nonumber \\ 
N&=&\frac{E^2}{2\pi \hbar c\sigma}+\ldots \ ,
\nonumber \\ 
S(E)&=&k_B \ln \Omega(N).
\label{SE2}
\end{eqnarray}
The microcanonical string entropy determines the temperature via 
\begin{eqnarray}
\frac{1}{T}=\frac{dS}{dE}
\ \ \ {\rm and}\ \ \ k_BT_H =\sqrt{\frac{3\hbar c \sigma }{4\pi }}\ ,
\nonumber \\ 
E=\left(\frac{T}{T-T_H}\right)\frac{7k_BT_H}{2} 
=\left(\frac{T}{T-T_H}\right)\Phi .
\label{SE3}
\end{eqnarray}
Eq.(\ref{SE3}) describes the thermodynamic excitation energy above the 
Hagedorn temperature $k_BT_H\approx 207\ {\rm MeV}$; i.e. 
$\lim_{T\to \pm \infty} E=\Phi \approx 725\ {\rm MeV}$. The string energy 
is plotted as a function of temperature in Figure\ref{fig1}.

The following is a simple physical picture for describing how QCD 
strings behave during a high energy nuclear collisions.
During a collision the meson (quark anti quark) or nucleon 
(three quark) string connections are excited. If a small amount of 
collision energy is transferred into a string, then the string temperature 
will be negative $0<E<\Phi \ \ \Rightarrow \ \ 0>T>-\infty $. If a large 
amount of collision energy is transferred into a string, then the string 
temperature will be positive $\Phi <E< \infty\ \ \Rightarrow \ \ T>T_H$ over 
and above the Hagedorn temperature. The string heat capacity 
\begin{equation}
C=\frac{dE}{dT}=-\frac{\Phi T_H}{(T-T_H)^2}<0 
\ \ \ \ \ \ {\rm (QCD\ string)}.
\label{SE4}
\end{equation}
The {\em color glass-like} time scales of string motions in the high energy 
limit are given in Eq.(\ref{intro4}). As $T\to T_H+0^+$, the time scales 
grow ever larger consistent with the VFT law often applied to polymer 
and other glasses. The QCD strings in nuclear matter thereby play the role 
of polymer (or other network) chains in melted glass with viscosity  
\begin{equation}
\eta =\rho c^2\tau
=\rho c^2\tilde{\tau} \exp\left[\frac{\Phi }{k_B(T-T_H)}\right],
\ \ \ (T>T_H),
\label{SE5}
\end{equation}
wherein $\rho $ is the fluid mass density. A color glass near the Hagedorn 
temperature is similar to a very viscous fluid. If the duration of a collision is 
sufficiently short, then there is little time to excite highly energetic 
string states leaving one with an intermediate temperature $T\gg T_H$ which 
implies a short time scale and a low viscosity. The resulting low viscosity 
in an almost {\em ideal fluid} is sometimes called 
a ``perfect fluid''\cite{Adcox:2005, Adams:2005}.

Finally, we note in passing that a black hole has thermodynamic analogies to a 
QCD string; Black holes, as well as some observed {\em normal stars}, also have 
negative heat capacity; e.g. for the black hole  
\begin{eqnarray}
\frac{S}{k_B}=4\pi \left(\frac{GM^2}{\hbar c}\right)
=4\pi \left(\frac{GE^2}{\hbar c^5}\right),
\nonumber \\ 
\frac{dE}{dS}=T\ \ \ \Rightarrow 
\ \ \ E=\left(\frac{\hbar c^5}{8\pi G}\right)\frac{1}{k_BT}
=\left(\frac{\hbar c\sigma_{\rm gravity}}{k_BT}\right),
\nonumber \\ 
C=\frac{dE}{dT}=
-\left(\frac{\hbar c\sigma_{\rm gravity}}{k_BT^2}\right)<0.
\label{BE}
\end{eqnarray}

\section{String Fragmentation \label{SF}}

What happens to a highly excited state QCD string when it decays into final 
product particles in a nuclear collision? If the string is highly excited, 
then what appears in the final state products are jets of the many parts 
of the strings which fragment into lower mass elementary particles.
To see what is involved, consider a meson made of a $q\bar{q}$ quark  
pair each with mass $m$ which sit on the ends of a QCD string. 
The quark pair Hamiltonian matrix may be written as 
\begin{equation}
H=\pmatrix{cp & mc^2 \cr
mc^2 & -cp}\ \ \ {\rm with\ \ eigenvalues}
\ \ \ E_\pm(p)=\pm\sqrt{c^2p^2+m^2c^4}.
\label{SF1}
\end{equation}
The transition rate per unit time to fragment a point on a string cereating 
a $q\bar{q}$ pair follows by applying Fermi's golden rule to the Hamiltonian 
matrix Eq.(\ref{SF1}); i.e. 
\begin{equation}
\gamma =\left(\frac{2\pi }{\hbar }\right)|mc^2|^2\delta (2cp).
\label{SF2}
\end{equation}
The probability of finding the fragmentation phase point $\in (dpdx)/2\pi \hbar $ is 
\begin{equation}
d^2P=\left(\frac{dpdx}{2\pi \hbar }\right) e^{-\int \gamma dt}\ .
\label{SF3}
\end{equation}
Since the rate of change of momentum of a quark in a color condensate 
electric field is given by 
\begin{equation}
\frac{dp}{dt}=g{\cal E}=\sigma ,
\label{SF4}
\end{equation}
it follows that the transition per unit time per unit length for a QCD string 
to fragment producing a $q\bar{q}$ pair is given by 
\begin{equation}
\frac{d^2 P}{dxdt}=\left(\frac{\sigma }{2\pi \hbar}\right)
\exp\left[-\frac{\pi m^2c^3}{\hbar \sigma }\right],
\label{SF5}
\end{equation}
which is exponentially suppressed if the quark mass $m$ is high.

\section{Conclusions}

It was shown how the perturbation theory QED vacuum is stable to decay  
but has a short space-like distance Landau ghost. The perturbation theory 
QCD vacuum has no Landau ghost, but is in an excited energetic state  
of negative temperature. From the viewpoint of QCD, nuclear matter consists 
of relativistic nucleons each with three quarks and relativistic 
mesons each with a quark anti quark pair. The quarks are tied together 
by gluons in the form of strings having tension $\sigma $. The excitation 
of QCD strings at low energy has a negative temperature and at high energy 
has a positive temperature always higher than $T_H$. Very high energy strings 
$T\to T_H+0^+$ move very slowly as a viscous melted glass with very high viscosity. 
However, in a very short collision time it is difficult to transfer the 
initial collision kinetic energy into the internal energy of a few strings. 
The low energy strings at high $T\gg T_H$ have a low viscosity giving rise 
to an almost perfect fluid. The role of strings in a nucleus is closely  
analogous to the role of polymer chains in some viscous glass beads.

\end{document}